\documentclass{ws-ijmpa}
\usepackage[super,compress]{cite}
\usepackage{graphicx}
\begin{document}
\markboth{Stanislav Iablokov}{Quantum Cherenkov radiation by spinless and spin-1/2 particles.}

%
\catchline{}{}{}{}{}
%

\title{Quantum Cherenkov radiation by spinless and spin-1/2 particles.}

\author{Stanislav Iablokov}

\address{Division of Theoretical Physics, Department of Physics,\\
Yaroslavl State P.\,G.~Demidov University, Sovietskaya 14,\\
150000 Yaroslavl, Russian Federation\\
iablokov@uniyar.ac.ru \footnote{\, The author could also be contacted via physics@iablokov.ru}}

\maketitle

\begin{history}
\received{Day Month Year}
\revised{Day Month Year}
\end{history}

\begin{abstract}
In this paper, the Cherenkov radiation process is investigated in the frame of the Quantum Field Theory, both for the spinless and spin-1/2 charged particles. In the latter case, the polarization density matrix technique was used to account for the spin flip. If compared with the Classical Electrodynamics results, in both cases several quantum corrections arise, coinciding up to the photon's energy linear term. It is shown that, while the absolute value of these corrections significantly relies on the refractive index, the relative quantum correction stays independent of the medium in the case of the ultrarelativistic particles. It the case of the spin-1/2 particles it turns out that the spin flip practically never occurs in this process. For both cases, spinless and spin-1/2, the outgoing photon polarization vector lies always in the plane formed by the initial particle's momentum vector and the final photon's wave vector.   

\end{abstract}

\ccode{PACS numbers: 41.60.Bq, 41.60.-m, 13.40.-f, 13.40.Hq, 13.35.-r, 13.88.+e}


\section{Introduction}	
Since its discovery, the Cherenkov radiation\cite{cherenkov,vavilov,tammfrank} became a very useful tool for particle registration and investigation of their properties. Given the fact that particles emit light only moving above some speed threshold, it became possible to distinguish between them experimentally. Nowadays, there exist a variety of the so called Cherenkov counters which are based on the effect.

In the very beginning of the Cherenkov radiation research, several papers\cite{jauchwatson,frank,ginzburg1,jelley,shwingertsaierber} were published concerning its quantum and polarization properties. And even at the present moment a strong interest to the effect still exists, resulting in the considerable amount of articles\cite{kaufholdklinkhamer,erberlatal,kharkov,chefranov,glue,weak} either devoted to this problem or concerning it from some point of view.

At the present moment, several quantum approaches to this phenomenon are known, although they all rely on the evaluation of the S-matrix element. Some of calculations are performed very explicitly and, therefore, the results have rather cumbersome form. In this paper, the calculations are made with use of the invariant amplitude technique, where the polarization density matrix\cite{landau4} approach is applied in the case of the spin-1/2 particles. This approach allows to obtain the intermediate and final expressions in a very compact form. The calculation results for the scalar field could be applied to the Cherenkov radiation of the $\pi^{\pm}$ mesons, while the results for the spinor field case could be applied to the Cherenkov radiation of any charged spin-1/2 particle.

A natural system of units is used, i.e., $c = \hbar = 1$, unless otherwise stated.

\section{Classical approach}

In the frame of Classical Electrodynamics, the Cherenkov radiation problem is solved by the calculation of the energy losses due to the medium polarization by a charged particle. In the most cases, see, e.g., Refs. \refcite{tammfrank,frank,landau7,jackson}, the analysis starts from the Maxwell equations in the medium treating a particle as a point-like source of electromagnetic field. Further steps require obtaining the expression for the effective force that causes energy losses due to the radiation. These techniques could be learned from the above mentioned references. There exist several ways to represent the final result: in terms of the effective force, in terms of the energy losses spectrum per unit time (or length), etc. The transition from one representation to another is trivial. In this article, it is found convenient to write the classical formula as follows:
\begin{equation}
\label{classical_formula}
dN = \frac{e^2}{n^2v}(n^2v^2-1)d\omega dt \, ,  
\end{equation}
where $dN$ is the number of photons emitted by the charged particle in the time interval $dt$ and in the angular frequency interval $d\omega$,  $v$ is the particle's velocity, $e$ is its charge, and $n$ is the refractive index of the medium.

This result is bounded by the condition:
\begin{equation}
\label{bound_speed}
v > \frac{1}{n}\,.
\end{equation}

Also the angle $\theta$ between the initial particle's momentum $\vec p$ and the photon's wave-vector $\vec k$ must satisfy the following relation:
\begin{equation}
\label{classical_angle}
\cos\theta =\frac{1}{nv}\,.
\end{equation}

Omitting the detailed derivation that could be found in classical papers or textbooks, it is worth mentioning the intuition that stays behind the expressions (\ref{bound_speed}) and (\ref{classical_angle}). When a particle moves at a velocity below the speed of light in the medium, the resulting field undergoes a destructive interference. In the superluminal case, the particle outstrips its wavefront that results the latter having a conic surface due to the constructive interference which we treat as an outgoing particle, i.e. a photon. For this to be valid, the relation (\ref{bound_speed}) should be satisfied. As for the expression (\ref{classical_angle}), it could be verified by the geometrical considerations.

\section{Quantum Field Theory approach}

According to the Quantum Field Theory, it is necessary to consider the $X \to X + \gamma$ process where $X$ is a charged particle, scalar or spinor. Due to the energy-momentum conservation law, this process is forbidden in vacuum. Another way of looking onto this limitation is the zero value of the reaction's phase volume:
\begin{equation}
\label{phase_volume}
\Phi = \int {\delta}^4(P-P'-q)\frac{d^3k}{\omega}\frac{d^3p'}{E'}\,,
\end{equation}
where $q^\mu = (\omega,\,\vec{k}\,)$,\,  $P^\mu = (E,\,\vec{p}\,)$ and ${P'}^\mu = (E',\,\vec{p}\,')$ are the 4-momenta of the photon, initial and final electrons, respectively. 

As opposed to the vacuum case ($n = 1$), in the medium ($n > 1$) one has:
\begin{equation}
\label{q_2}
q^2 = \Pi(\omega) = -\omega^2(n^2-1) \ne 0 ,
\end{equation}

where $\Pi(\omega)$ is the polarization operator's eigenvalue\footnote{\,See Eq. (43.52) in [\citen{akhiezer}]}.

In this case, $\Phi$ is not zero and the reaction becomes possible. From the energy-momentum conservation law, it follows:
\begin{equation}
\label{quantum_angle}
\cos\theta = \frac{1}{nv} \left\{ 1+(n^2-1) \frac{\omega}{2E} \right\}\, ,
\end{equation}
thus leading to a slight decrease of the angle if compared with (\ref{classical_angle}).

In order to calculate the number of the Cherenkov photons emitted per second, the conventional method is applied. It relies on the notion of the S-matrix element:
\begin{eqnarray}
\label{smatrix1}
{\cal S}_{X \to X\gamma} =\,< X\gamma\,| \, \hat{T} e^{\mathrm{i} \int \mathrm{d}^4 x\, \hat{\cal L}}  |\,X >\, .
\end{eqnarray}

Extracting by the standard procedure the invariant amplitude ${\cal M}_{X \to X\gamma}$, one arrives to the differential (per unit time) probability of the process by the integration over the reaction's phase volume:
\begin{eqnarray}
\label{width1}
\Gamma = \frac{\mathrm{d}{\cal W}}{\mathrm{d}t} = \int (2\pi)^4 \delta^4(P-P'-q)\frac{|{\cal M}_{X \to X\gamma}|^2 \mathrm{d}^3 \vec{p'} \, \mathrm{d}^3 \vec{k}}{2E 2E' (2\pi)^3 2\omega(2\pi)^3} \, .
\end{eqnarray}
It can be treated as the amount of photons emitted per second by a charged particle through the Cherenkov mechanism.
Having $\Gamma$ represented in the form
\begin{equation}
\label{width2}
\Gamma = \int \frac{\mathrm{d}\Gamma}{\mathrm{d}\omega} \mathrm{d}\omega \, ,
\end{equation}
the spectrum for $\Gamma$, $\mathrm{d}\Gamma / \mathrm{d}\omega$, i.e. the number of emitted photons per unit time per interval of the angular frequency $\mathrm{d}\omega$, could be extracted from (\ref{width1}).

In the original expression (\ref{width1}), the integration is made over $\mathrm{d}^3 \vec{k}$, and therefore $\mathrm{d}k$ appears in the final answer. The desirable form (\ref{width2}) could be obtained by the next procedure.

In the medium, there arises a charge shielding, so it is necessary to account for radiative corrections. This leads to the following substitution: 
\begin{equation}
\label{renormalization}
e \to e_{shielded} = e\sqrt{Z}\,,
\end{equation}
where the renormalization factor has the form:
\begin{equation}
\label{renormalization_factor}
Z = \frac{1}{1-\frac{d\Pi (\omega)}{d(\omega ^2)}} = \frac{1}{n} \frac{d\omega}{\mathrm{d}k}\,.
\end{equation}

Having $|{\cal M}_{X \to X\gamma}|^2 \sim e^2$ and making the substitution (\ref{renormalization}), one arrives to the following rule:
\begin{equation}
\label{renormalization_rule}
|{\cal M}_{X \to X\gamma}|^2 \to {|{\cal M}_{X \to X\gamma}|}^2_{shielded} = \frac{1}{n} \frac{d\omega}{\mathrm{d}k} {|{\cal M}_{X \to X\gamma}|}^2 \, .
\end{equation}

The integral in (\ref{width1}) hence takes the form:
\begin{eqnarray}
\label{width4}
\Gamma =  \int \frac{(2\pi)^4 \delta^4(P-P'-q)|{\cal M}_{X \to X\gamma}|^2}{2E 2E' (2\pi)^3 2\omega(2\pi)^3} \frac{1}{n} \frac{d\omega}{\mathrm{d}k} \mathrm{d}^3 \vec{p'} \, \mathrm{d}^3 \vec{k} \, .
\end{eqnarray}

It might look like the radiative correction would complicate the analytical evaluation of the integral, but in fact it leads to the simple transition from $\mathrm{d}k$ to $\mathrm{d}\omega$:
\begin{eqnarray}
\label{dkdomega}
\frac{\mathrm{d}\omega}{\mathrm{d}k} \mathrm{d}k = \mathrm{d}\omega\,.
\end{eqnarray}


\section{Spinless and spin-1/2 particles}

Let's consider the cases of scalar and spinor fields. The explicit forms of the respective field operators, scalar ($\phi$), spinor ($\psi$) and electromagnetic ($A^\mu$) are:
\begin{eqnarray}
\label{scalar_field}
{\phi}(x) = \sum\limits_{\vec{p}} \frac{1}{\sqrt{2E_{\vec{p}}V}}\left[ a_{\vec{p}} \, e^{-i(px)} + b_{\vec{p}}^+ \, e^{+i(px)} \right] \, ,
\end{eqnarray}

\begin{eqnarray}
\label{spinor_field}
{\psi}(x) = \sum\limits_{\vec{p},s} \frac{1}{\sqrt{2E_{\vec{p}}V}}\left[ a_{\vec{p}, s} \,u_{(s)}(p) e^{-i(px)} + b_{\vec{p}, s}^+ \, u_{(s)}(-p) \, e^{+i(px)} \right] \, ,
\end{eqnarray}

\begin{eqnarray}
\label{em_field}
A^{\mu}(x) = \sum\limits_{\vec{k},\lambda} \frac{\sqrt{4\pi}}{\sqrt{2\omega_{\vec{k}} V}}\left[ c_{\vec{k}, \lambda} {\varepsilon}_{(\lambda)}^{\mu} e^{-i(kx)} + c_{\vec{k}, \lambda}^+ \, {\varepsilon}_{(\lambda)}^{* \mu} \, e^{+i(kx)} \right] \, .
\end{eqnarray}

Here, $u_{(s)}$ are the bispinors corresponding to the possible particle's polarizations ($s = \pm 1$), ${\varepsilon}_{(\lambda)}^{\mu}$ are the electromagnetic polarization vectors. The respective creation and annihilation operators of the fields are denoted as $a / a^+$, $b / b^+$ and $c / c^+$. In this paper, the following gauge\footnote{\,The results obtained in this gauge coincide with the results obtained in other gauges, see, e.g., Ref. \refcite{erberlatal} where all three electromagnetic polarizations were considered} is chosen for $A^\mu$:
\begin{eqnarray}
\label{em_gauge}
A^{\mu} = (A^0, \vec{A})\,, \quad A^0 = 0 \, , \quad \mathrm{div} \vec{A} = 0.
\end{eqnarray}

In this gauge, only two polarizations are present with their vectors being orthogonal to each other and to the photon's wave-vector. They are also taken to be real, i.e.:
\begin{eqnarray}
\label{polarization_vectors}
{\varepsilon}_{(\lambda)}^{\mu}k_\mu = 0, \quad {\varepsilon}_{(\lambda)}^{\mu} = {\varepsilon}_{(\lambda)}^{* \mu}  \quad ({\lambda=1,2}) \, .
\end{eqnarray}
%
For further integration purposes, the {\it z}\,-axis is chosen to be along $\vec{p}$, {\it x}\, and {\it y}\, axes could be chosen arbitrarily. This leads to the following expressions:
\begin{eqnarray}
\label{kp_coords}
k^\mu = (\omega, k \sin\theta \cos\phi, k \sin\theta \sin\phi, k\cos\theta)\, , \quad p^\mu = (E, 0, 0, p )\, ,
\end{eqnarray}
where $\theta$ and $\phi$ are the corresponding spherical angles.

To analyse the final results for the partial amplitudes it is convenient to choose the polarization vectors as follows:
\begin{eqnarray}
\label{polarization_coords}
\nonumber
{\varepsilon}_{(1)}^{\mu} = (0, \vec{\varepsilon}_{(1)}) = (0, \sin\phi, -\cos\phi, 0)\,, \quad \\
{\varepsilon}_{(2)}^{\mu} = (0, \vec{\varepsilon}_{(2)}) = (0, \cos\theta \cos\phi, \cos\theta \sin\phi, -\sin\theta).
\end{eqnarray}

Throughout the calculations the following notations are used: $\gamma^5 = \mathrm{i}\gamma^0\gamma^1\gamma^2\gamma^3$, ``hatted'' vector $\hat{a} = a_\mu\gamma^\mu$ and $(ab)=a_\mu b^\mu$.

For the spinor case the notion of the polarization density matrix is used, both, partial:
\begin{eqnarray}
\label{density_matrix_partial}
\rho^{(s)}(p) = u^{(s)}(p) \bar{u}^{(s)}(p) = \frac{1}{2} (\hat{p} + m)(1 + \gamma_5\hat{s}),
\end{eqnarray}
and summed over the possible fermion's polarizations: 
\begin{eqnarray}
\label{density_matrix_summed}
\rho(p) = \sum\limits_{s} u_{(s)}(p) \bar{u}_{(s)}(p) = (\hat{p} + m).
\end{eqnarray}

Here, $s^\mu$ is the 4-dimensional polarization vector which characterizes the fermion's spin properties. In the rest frame it has the following form: 
\begin{eqnarray}
\label{4spin_restframe}
s^\mu = (0, \vec{\zeta})\, ,
\end{eqnarray}
where $\vec{\zeta}$ is twice the mean spin vector in the fermion's rest frame. Under the Lorentz boost it transforms to:
\begin{eqnarray}
\label{4spin_lorentz}
s^\mu = \bigg( \frac{(\vec{\zeta}\vec{p})}{m} , \vec{\zeta} + \frac{\vec{p}(\vec{\zeta}\vec{p})}{m(E+m)}\bigg)\, .
\end{eqnarray}

Using these formulas, one could perform calculations of the corresponding matrix elements for the cases of the scalar and spinor particles. 
\subsection{Scalar field}

The Lagrangian of the electromagnetic interaction for the scalar field has the form:
\begin{equation}
\label{scalar_lagrangian}
{\cal L}^{int} = - ie(\phi^* \overset{\leftrightarrow}{\partial _{\mu}} \phi)A^{\mu}\, .
\end{equation}
For a partial invariant amplitude of the process $X \to X + \gamma$ one obtains:
\begin{eqnarray}
\label{scalar_amplitude}
{\cal M}_{X \to X\gamma}^{(\lambda)} = - \sqrt{4\pi}e (p^\mu+{p'}^{\mu}){\varepsilon}^{(\lambda)}_\mu = - 2\sqrt{4\pi}e (p{\varepsilon}^{(\lambda)})\,,
\end{eqnarray}
\begin{eqnarray}
\label{scalar_amplitude2}
|{\cal M}_{X \to X\gamma}^{(\lambda)}|^2 = 16\pi e^2 (p{\varepsilon}^{(\lambda)})^2\,.
\end{eqnarray}
Summation over the photon polarizations results in:
\begin{equation}
\label{scalar_amplitude2summed}
|{\cal M}_{X \to X\gamma}|^2 = \sum\limits_{\lambda} |{\cal M}_{X \to X\gamma}^{(\lambda)}|^2 = 16\pi e^2 {\vec{p}}^{\,\,2} \sin^2\theta\,.
\end{equation}
Performing partial integration in \eqref{width4} one obtains the following formula for the photon spectrum in the case of the scalar charged particle:
\begin{equation}
\label{scalar_spectrum}
\frac{\mathrm{d}\Gamma}{\mathrm{d}\omega}  = \frac{e^2}{n^2v}\left( n^2v^2-1 - \frac{\omega (n^2-1)}{E} - \frac{\omega^2(n^2-1)^2}{4E^2} \right)\,.
\end{equation}


\subsection{Spinor field}

In the case of the spinor Dirac field, the interaction Lagrangian could be written as:
\begin{eqnarray}
\label{spinor_lagrangian}
{\cal L}^{int} = -e\bar{\psi}\hat{A}\psi\,.
\end{eqnarray}
The corresponding partial invariant amplitude would look as follows:
\begin{eqnarray}
\label{spinor_amplitude}
{\cal M}_{X \to X\gamma}^{(s)(s')(\lambda)} = - \sqrt{4\pi} e\bar{u}^{(s')}(p')\hat{\varepsilon} ^ {(\lambda)}u^{(s)}(p) \, .
\end{eqnarray}


First, let's sum the square of (\ref{spinor_amplitude}) over the final fermion's polarizations and average over the initial ones:
\begin{eqnarray}
\label{spinor_amplitude2_summed_ss}
|{\cal M}_{X \to X\gamma}^{(\lambda)}|^2 = \frac{1}{2} \sum\limits_{s, s'} |{\cal M}_{X \to X\gamma}^{(s)(s')(\lambda)}|^2 = 16\pi e^2 \left[ (p{\varepsilon}^{(\lambda)})^2 -\frac{q^2}{4} \right] \, .
\end{eqnarray}
Summation of (\ref{spinor_amplitude2_summed_ss}) over the photon polarizations results in:
\begin{eqnarray}
\label{spinor_amplitude2_summed_sslambda}
\left|{\cal M}_{X \to X\gamma}\right|^2\ = \sum\limits_{\lambda} \left|{\cal M}_{X \to X\gamma}^{(\lambda)}\right|^2 = 16\pi e^2 \left( {\vec{p}}^{\,\,2} \sin^2\theta  - \frac{q^2}{2} \right) \, .
\end{eqnarray}
Integration over the phase space leads to final formula for the spectrum:
\begin{equation}
\label{spinor_spectrum}
\frac{\mathrm{d}\Gamma}{\mathrm{d}\omega}  = \frac{e^2}{n^2v}\left( n^2v^2-1 - \frac{\omega (n^2-1)}{E} + \frac{\omega^2(n^4-1)}{4E^2} \right) \, .
\end{equation}

For further analysis, it is also interesting to estimate the spin flip probability by the consideration of the corresponding partial amplitudes. Starting with the square of (\ref{spinor_amplitude}) and the explicit form (\ref{density_matrix_partial}) of the partial polarization density matrix, one arrives to:
\begin{eqnarray}
\label{spinor_amplitude2_partial3}
\nonumber
|{\cal M}_{X \to X\gamma}^{(s)(s')(\lambda)}|^2 = 8\pi e^2 \bigg \lbrace \left[(p{\varepsilon}^{(\lambda)})^2 -\frac{q^2}{4}\right]\left[1-(ss')\right] + \frac{q^2}{2}({\varepsilon}^{(\lambda)}s)({\varepsilon}^{(\lambda)}s') + \\
+ ({\varepsilon}^{(\lambda)}p)({\varepsilon}^{(\lambda)}s)(qs') - ({\varepsilon}^{(\lambda)}p)({\varepsilon}^{(\lambda)}s')(qs) -\frac{1}{2}(qs')(qs) \bigg\rbrace \, . \quad
\end{eqnarray}
Summing (\ref{spinor_amplitude2_partial3}) over the photon polarizations, one gets:
\begin{eqnarray}
\label{spinor_amplitude2_partial4}
|{\cal M}_{X \to X\gamma}^{(s)(s')}|^2 = \sum\limits_{\lambda} |{\cal M}_{X \to X\gamma}^{(s)(s')(\lambda)}|^2 = 8\pi e^2 \bigg\lbrace {\vec{p}}^{\,\,2} \sin^2\theta\left[1-(ss')\right] + \alpha (E\omega) \bigg\rbrace \, , \quad
\end{eqnarray}
where the $\omega/E$ ratio is taken to be a small parameter, as it will be discussed later, and therefore $\alpha (E\omega)$ denotes the expression of order $E\omega$ or $\omega^2$.


\section{Analysis}

Let's analyze the obtained results. Expressions (\ref{scalar_spectrum}) and (\ref{spinor_spectrum}) coincide up to the $\omega/E$ linear term. They differ in quadratic terms due to the spinor nature of the fermion field. For the typical values of the parameters ($E \sim 10MeV$, $\omega \sim 1eV$, $n \sim 1$) these terms are negligible if compared with the linear one, and therefore one could conclude that {\it irrespective of the field type, scalar or spinor, the quantum correction is the same.} The resulting formula
\begin{equation}
\label{final_formula}
\frac{\mathrm{d}^2N}{\mathrm{d}t\mathrm{d}\omega}  = \frac{\mathrm{d}\Gamma}{\mathrm{d}\omega}  = \frac{e^2}{n^2v}\left( n^2v^2-1 - \frac{\omega (n^2-1)}{E} \right)
\end{equation}
is in a good accordance with the existing results and, as expected, turns into the classical one, Eq. (\ref{classical_formula}), in the limit of low frequencies.

A relative correction to the number of emitted photons predicted by the classical formula (\ref{classical_formula}) is given by
\begin{equation}
\label{correction1}
\frac{\delta N}{N} = - \frac{\omega (n^2-1)}{E(n^2v^2-1)}\,, 
\end{equation}
thus being dependent on the medium through the refractive index $n$.
But in the very special case when $v$ tends to $1$, the expression (\ref{final_formula}) turns into
\begin{equation}
\label{correction2}
\frac{\mathrm{d}^2N}{\mathrm{d}t\mathrm{d}\omega}  = \frac{\mathrm{d}\Gamma}{\mathrm{d}\omega}  = \frac{e^2}{n^2v}(n^2-1)\left(1  - \frac{\omega}{E} \right)\,,
\end{equation}
therefore leading to the simplification of (\ref{correction1}):
\begin{equation}
\label{correction3}
\frac{\delta N}{N} = - \frac{\omega}{E} \, . 
\end{equation}
Obviously, {\it in the case when $v$ tends to $1$ the relative quantum correction is independent of the medium.}
This fact could have some use in the precise measurements of quantum corrections to the Cherenkov radiation if the corresponding techniques would be improved. At the present moment it is less likely to measure such a small divergence.

Next, one would like to consider the polarization properties of the emitted light. Let's call the plane where $\vec{p}$, $\vec{p}\,'$ and $\vec{k}$ vectors lie as the plane of propagation. From (\ref{scalar_amplitude2}), it follows that for the scalar field all radiation would have the polarization vector lying in that plane because of the partial amplitude for $\lambda = 1$ which is equal to zero. That wouldn't be so for the spinor field. In (\ref{spinor_amplitude2_summed_ss}) there is a small correction term $-q^2 / 4 > 0$, hence resulting in additional radiation with the polarization vector having a component perpendicular to the  plane of propagation. But as far as the order of magnitudes in this problem ($E \sim 10MeV$, $\omega \sim 1eV$, $n \sim 1$) is such as it's been previously mentioned, this term could be neglected. Therefore, {\it for both cases, scalar and spinor, the emitted radiation is polarized in the plane of propagation}.

Another interesting feature follows from the spin flip consideration. In the final formula for the partial amplitudes summed over the photons' polarizations (\ref{spinor_amplitude2_partial4}) there are terms of order $\omega^2$ and $E\omega$. They are negligible if compared with the $E^2$ term, and therefore could be omitted. Because of the relatively low value of the photon's energy and momentum the initial and final values of the electrons' energy and momentum could be taken as equal. From this, one could notice that it is possible to write $s^\mu$ and $s'^\mu$ as follows:
\begin{eqnarray}
\label{ssprime}
s^\mu = \tau \sigma^\mu = \tau \bigg(\frac{(\vec{p}\,\vec{\zeta})}{m}, \vec{\zeta} + \frac{\vec{p}(\vec{p}\,\vec{\zeta})}{m(E+m)}\bigg)\, , \quad s'^\mu = \tau ' \sigma^\mu \, .
\end{eqnarray}

As long as $\sigma^\mu\sigma_\mu = -1$, expression (\ref{spinor_amplitude2_partial4}) simplifies to
\begin{eqnarray}
\label{spin_flip}
|{\cal M}_{e \to e\gamma}^{(\tau)(\tau')}|^2 = 8\pi e^2 \bigg \lbrace {\vec{p}}^{\,\,2} sin^2(\theta) (1+\tau\,\tau')  \bigg\rbrace\, ,
\end{eqnarray}
thus leading to the conclusion, that {\it the radiation with antiparallel (i.e. $\tau\tau' = -1$) spin orientations is highly suppressed, while the conservation of spin projection to the $\vec{\zeta}$\,-axis is promoted}. 

\section{Conclusions}

To the best knowledge of the author, the statement about the coincidence of the quantum corrections for the scalar and spinor fields hasn't been previously mentioned in the papers concerning the quantum properties of the Cherenkov radiation. The medium independence of the relative quantum correction in the case when $v$ tends to $1$, despite its evidence, also hasn't been explicitly pointed out in the literature. 

Polarization properties of the radiation were established experimentally by P.Cherenkov himself and were presented in the original {paper\cite{cherenkov}}, their theoretical treatment in the frame of Classical Electrodynamics also exists. In this paper, it was found convenient to write the polarization conditions (\ref{scalar_amplitude2}) and (\ref{spinor_amplitude2_summed_ss}) in the explicitly covariant form in terms of the partial amplitudes of the considered process, also showing that they apply both for the scalar and spinor cases. Speaking about the spin flip, it is worth saying that this fact was first theoretically established by Ginzburg\cite{ginzburg1} who studied the change in what we now call as helicity. In this paper, the ``spin flip'' condition is investigated by the means of the polarization density matrix, therefore not limiting the spin axis to the direction of propagation.

Being a very powerful tool for the particle registration and investigation of their properties, the Cherenkov radiation effect is still open for new ideas. The purpose of this work was to fill the gaps concerning scalar fields and relative quantum corrections, at the same time providing the convenient framework for the polarization and spin flip considerations based on the notion of the polarization density matrix. The results obtained in this paper, being purely theoretical at the moment, might have practical use in the future in case of the measurement technique improvement.

\newpage
\section*{Acknowledgments}

This paper is dedicated to the memory of Nikolay Vladimirovich Mikheev who originally proposed the idea to study the Cherenkov radiation of the scalar particles and quantum corrections to it. Also, the author would like to thank A.V.Kuznetsov for his valuable comments and corrections.	


\end{document}